\documentclass[%
 aip,
 sd,%
 amsmath,amssymb,
preprint,%
]{revtex4-1}

\usepackage{graphicx}
\usepackage{dcolumn}
\usepackage{bm}
\usepackage{color}

\begin{document}

\preprint{  }

\title{Effects of beam velocity and density on an ion-beam pulse moving in magnetized plasmas}

\author{Xiao-ying Zhao}
\affiliation{Institute of Modern Physics, Chinese Academy of
Sciences, Lanzhou 730000, China}
\author{Hong-Peng Xu}%
\affiliation{Henan Polytechnic Institute, Nanyang 473000, China}

\author{Yong-tao Zhao}
\affiliation{School of Science, Xi'an Jiaotong University, Xi'an 710049, China}
\author{Xin Qi}
\thanks{Corresponding author:qixin2002@impcas.ac.cn}
\affiliation{Institute of Modern
Physics, Chinese Academy of Sciences, Lanzhou 730000, China}%
\author{Lei Yang}
\thanks{Corresponding author:lyang@impcas.ac.cn}
\affiliation{Institute of Modern
Physics, Chinese Academy of Sciences, Lanzhou 730000, China}%

\date{\today}

\begin{abstract}

The wakefield and stopping power of an ion-beam pulse moving in magnetized plasmas are investigated by
particle-in-cell (PIC) simulations. The effects of beam velocity and density on the wake and stopping power are discussed.
In the presence of magnetic field, it is found that beside the longitudinal conversed V-shaped wakes, the strong whistler wave are observed when low-density and low-velocity pulses moving in plasmas. The corresponding stopping powers are enhanced due to the drag of these whistler waves. As beam velocities increase, the whistler waves disappear, and only are conversed V-shape wakes observed. The corresponding stopping powers are reduced compared with these in isotropic plasmas. When high-density pulses transport in the magnetized plasmas, the whistler waves are greatly inhibited for low-velocity pulses and disappear for high-velocity pulses. Additionally, the magnetic field reduces the stopping powers for all high-density cases.

\end{abstract}

\maketitle

\section{\label{sec:level1}INTRODUCTION}

Understanding the interactions of charged particles with magnetized plasmas has been an interesting topic for several decades.
The stopping power of ion beams as well as plasma waves excited by the charged particles are important for fundamental physics\cite{Drake} and many applications such as electron cooling of ion beams\cite{Sorensen1983,Goldman1990}, inertial confinement fusion (ICF) driven by ion beams\cite{Renk2008,Ter2008} and neutral beam injection (NBI) in magnetically confined fusion plasmas\cite{Thompson1993,Takahashi2004}.

When the charged particles inject into the plasmas, the wakefield exhibits characteristic structures due to the response of
plasma electrons to the disturbance from the charged particles\cite{wake1,wake2,wake3,wake4,wake5,wake6}.
The structure of the wakefield is important in plasma wake-field accelerator\cite{acceler,Caldwell2009}, where the wakefield induced by ion beams can be used to accelerate the charged particles. In the case of isotropic plasmas, I. D. Kaganovich \emph{et al.} developed the analytical electron fluid model to describe the plasma response to a propagating ion beam\cite{wake4}. They showed that the ion beam would be neutralized by plasmas and formed a V-shape cone structure in the wake-field region. However, when ion beams moving in a magnetized plasma, the wakes will be more complicated.
The wakefield is found to be highly asymmetrical when a strong magnetic field is imposed perpendicular to the trajectory of a test charged particle\cite{wake5}. For ion beams moving parallel to the external magnetic field, the wakes will lose their typical V-shape cone structures and exhibits conversed V-shape structures\cite{wake2}. Specially, M. A. Dorf \emph{et al.} carried out a linear theoretical analysis and demonstrated that electromagnetic wave-field perturbations propagating oblique to the beam axis will be excited when the magnetic field strength satisfies $\omega_{ce}>2\beta_b \omega_{pe}$ and the ion beam pulse is sufficiently long $l_b\gg V_b / \omega_{pe}$\cite{Milhail2010}. Here, $\omega_{ce}$ and $\omega_{pe}$ are the electron cyclotron frequency and electron plasma frequency, respectively. $\beta_b=V_b/c$ is the ion beam velocity normalized to the speed of light and $l_b$ is the length of ion beam pulse.

Another important quantity of ions moving in magnetized plasmas is the stopping power, which is defined as the energy change per unit path-length $dE/ds$. When the pulses transport in magnetized plasmas, the movement of plasma electrons will be restricted in the magnetic field direction. The influence of the magnetic field on the nonlinear stopping power in plasmas was widely investigated for many years\cite{Boine1996,Zwicknagel1998,Gericke1999,Hu2009,Zwicknagel1996,Bringa1995,Walter2000,Mollers2003}.
For low ion velocity$(V_{b0}<10 V_{th})$, an enhancement of the stopping power for ions moving transversal to the magnetic field was found\cite{Walter2000} in a electron plasma, where $V_{th}=\sqrt{k_BT_e/m_e}$ is the thermal velocity of plasma electrons. And if the ion moves parallel to the magnetic field, it is found that the energy loss is reduced\cite{Mollers2003}. Moreover, the collective effect\cite{wake1, collective1,collective2} plays important roles for high-density beam pulses transport in magnetized plasmas.  It is reported  that strengthening the magnetic field reduces the stopping power for low- and high-density pulses. In regions of  moderate beam density, the stopping power increases in a weak magnetic field, but decreases in a strong magnetic field\cite{wake2}.

 In this paper,  2D3V-PIC simulations are performed to investigate the process of a hydrogen ion-beam pulse moving longitudinal through hydrogen magnetized two-component plasmas. The effect of ion-beam pulse with different velocities and densities on the wakefield and stopping power are studied. All simulations are performed using the code VORPAL\cite{vorpal}.
The paper is organized as follow: In Sec. II, the PIC simulation methods used in the paper are briefly described.
In Sec. III, the effects of beam velocity and density on the wake field and stopping power are obtained and discussed. The  summary is given in Sec. IV.

\section{\label{sec:level2}PARTICLE-IN-CELL SIMULATION METHODS}

We regard the plasmas as an assembly of charged particles, considering the two-dimensional (2D) plasma slab model shown in
Figure 1. The external magnetic field $B$ applied in the plasmas is uniform and directed along the x axis.
The dimensions of the simulation region are $x=0$ to $x=L_x$ and $y=0$ to $y=L_y$.
Initially, the plasmas (including plasma electrons and ions) are placed in the box, and the ion-beam pulse is placed on the left side of the simulation box and assumed to be moving in the x direction with an initial velocity $V_b$. The simulation uses a 2D3V electromagnetic PIC code.
All the charged particles are considered to move on the $x-y$ plane.

The equations of motion for all particles involved in the simulations
(plasma electrons, ions, and the injection-beam ions) are:

\begin{equation}\label{eq:1}
  \frac{d \mathbf{r}_j^{\alpha}}{d t}=\mathbf{v}_j^{\alpha},
\end{equation}

\begin{equation}\label{eq:5}
  \frac{d (\gamma_j^{\alpha} \mathbf{v}_j^{\alpha})}{d t}=\frac{q_{\alpha}}{m_{\alpha}}(\mathbf{E}+\mathbf{v}_j^{\alpha} \times \mathbf{B}),      j=1,2,...,N_\alpha.
\end{equation}

Here, $\mathbf{r}_j^{\alpha}$, $\mathbf{v}_j^{\alpha}$,
$q_{\alpha}$, $\gamma_j^{\alpha}$, $m_{\alpha}$, and $N_\alpha$ are
the position, velocity, charge, Lorentz factor, mass, and total
number of plasma electrons ($\alpha=e$), ions ($\alpha=i$), and
injection-beam ions ($\alpha=b$), respectively. The electric and
magnetic fields are updated by the Faraday's equation and the
Ampere-Maxwell equation,

\begin{equation}\label{eq:3}
  \frac{\partial \mathbf{B}}{\partial t}=-\nabla \times \mathbf{E},
\end{equation}

\begin{equation}\label{eq:4}
  \frac{\partial \mathbf{E}}{\partial t}=c^{2}\nabla \times \mathbf{B}-\frac{\mathbf{J}}{\varepsilon_{0}},
\end{equation}
where the particle source is

\begin{equation}\label{eq:5}
  \mathbf{J}=\sum_j q_{\alpha} \mathbf{v}_j^{\alpha} \delta(x-x_j).
\end{equation}

All simulations were performed by VORPAL software. The simulation box is composed of 800 grids on the x axis and 512 grids
on the y axis. The space and time steps are chosen to be $dx=dy=\lambda_e=5.25 \times 10^{-5}m$ and $dt=9.1 \times 10^{-14}s$, respectively, to satisfy
 the Courant-Friedrichs-Lewy (CFL) limit, where $\lambda_e=\sqrt{\varepsilon_{0}K_{B}T_{e0}/N_{e0}e^{2}}$ is the Debye length of plasma electrons.
50 superparticles are placed per cell in simulations.
A periodic boundary condition in the y direction and an open boundary condition in the x direction are adopted. The plasmas are composed of electrons and
H ions. A uniform plasma density $N_{e0}=N_{i0}=2.0 \times 10^{17}m^{-3}$ and an initial plasma temperature $T_{e0}=T_{i0}=10eV$ are used
for electrons and ions, respectively. The beam ions are chosen to be protons with a Gaussian shape distribution:
$N_b(x,y)=\rho_{b0}exp(-\frac{(x-x_0)^2}{(L_{b}/2)^2})exp(-\frac{(y-y_0)^2}{R_{b}^2})$,
where the length $L_{b}=40\times \lambda_e$ and radius $R_{b}=2
\times \lambda_e$. All particles are considered to be charged robs
moving on the x-y plane and the simulation tracks the following coordinates for each particle: $x, y, V_x, V_y, V_z$, which goes partway to a 3D model.
At each time step, the energy loss $\Delta E$ and the travel path $\Delta s$ of every injection-ion are recorded. The
stopping power per ion is calculated by averaging $\Delta E
/ \Delta s $ over the entire simulation and over all particles.

\section{\label{sec:level3}Simulation results and discussions}

The influence of beam velocity on the wakefield and stopping power when beam density is smaller than electron density $\rho_{b0}=1.0 \times 10^{17} m^{-3}$(the linear case) is first investigated. Figure 2 shows the evolution of the electric wakefield induced by the ion-beam pulse moving in a magnetic plasma when beam velocity $V_b=0.025c$ (satisfy $L_b\gg V_b /\omega_{pe}$). At the very beginning, as shown in Figure 2(a), the whistler wave-field perturbations propagating oblique to the beam axis is observed, which is consistent with the linear theoretical analysis prediction by M. A. Dorf \emph{et al.}\cite{Milhail2010}. When T=1.4ns as shown in Figure 2(b), the wakes exhibit whistler waves, while there is no obvious structure behind the trajectory. In Figure 2(c) when T=2.1ns, besides the whistler waves, the conversed V-shape cone structures are observed. In order to further investigate the influence of the magnetic field on the wakefield, Figure 3 shows the comparison of the electric wakefield induced by the ion-beam pulse moving in (a), isotropic plasmas and (b), magnetized plasmas. As $B=0.0T$ in Figure 3(a), the wakes exhibit the fluctuation behind the pulse. However, as $B=1.0T$ in Figure 3(b), the wakes show both conversed V-shaped cone and whistler wave structures. Moreover, the amplitude of the whistler waves is higher than that of the conversed V-shape fluctuation. The corresponding longitudinal electric field along the trajectory of the pulse for isotropic and magnetic cases is shown in Figure 3(c). For low-density and low-velocity beam pulses, the length of ion-beam pulse is much greater than the wavelength of electron plasma wave excitations, $L_b\gg V_b /\omega_{pe}$. Therefore, electrostatic electron plasma wave excitations are significantly suppressed and there is no obvious wave just behind the ion-beam pulse for both cases.

For higher velocity when $V_{b0}=0.15c$ ($L_b\approx V_b /\omega_{pe}$), results are shown in Figure 4. Figure 4(a) is the wakefield in isotopic plasmas and Figure 4(b) is the result in magnetized plasmas.When $B = 1.0T$(Figure 4(b)), the conversed V-shape cone structures are observed. Comparing with  the structures in Figure 3(b), the whistler waves disappear in the wake region. This is due to that for beam pulses with high velocity, the electrostatic electron plasma wave excitations play dominant roles. Figure 4(c) shows the longitudinal electric field along the trajectory of both cases. It can be seen clearly that in the presence of the magnetic field, the wakefield is inhibited. This is due to that plasma electrons are restricted in the direction perpendicular to the magnetic field.

Figure 5 shows the dependence of the stopping power per pulse ion on the beam velocity with different magnetic field $B = 0.0T$ and $B = 1.0T$. It is noted that for low-velocity beams ($V_{b0}<0.05c$), the presence of magnetic field enhances the stopping power. This is because that in the presence of the magnetic field, whistler waves are excited by the ion-beam pulse, as shown in Figure 3(b). These wakes draw back the beam and enhance the stopping power. As the velocity increases, ($V_{b0}>0.05c$), the stopping power is reduced in the presence of the magnetic field.

The above discussion mainly concerns conditions when the density of the pulses is smaller than the electron density in plasmas. When the density of beam pulses is much more higher than the density of the electron, the collective effects and screening electrons play important roles.
Figure 6 shows the electric wakefield induced by the ion-beam pulse in case of isotropic plasma and magnetic plasma
when $\rho_{b0}=2.0 \times 10^{18} m^{-3}$ and $V_{b0}= 0.025c$ . At $B=0.0T$ in Figure 6(a), the wakes exhibit typical V-shaped structures behind the pulse. At $B=1.0T$ in Figure 6(b), the whistler waves are observed. However, comparing with Figure 3(b), the amplitude of the whistler waves is obviously reduced. This is due to that for high-density pulses transport in plasmas, nonlinear perturbations caused by collective effects weaken the whistler wave excitations. The longitudinal electric field along the trajectory is shown in Figure 6(c). As discussed above, the wakefield is inhibited in the presence of the magnetic field. For higher velocity when  $V_{b0}=0.15c$, results are given in Figure 7. The wakes exhibit the apparent V-shape cone structures when $B=0.0T$, as shown in Figure 7(a). The conversed V-shaped cone structures appear when $B=1.0T$ as shown in Figure 7(b). Moreover, as the velocity increases, the whistler waves disappear, either. The longitudinal electric fields on the trajectory  are given in Figure 7(c). It can be seen clearly that in the presence of the magnetic field, the amplitude of the wakefield becomes smaller. This is because in the presence of magnetic field, more electrons are restricted in the transverse direction and cannot response actively to the attraction of the pulses. Moreover, The magnetic field reduces the screening of the ion-beam pulses. As the screening effect of the electrons weakens, the collective effect of the ion-beam pulse becomes more significant, which enhances the electron trapping process and inhibits the fluctuation of the wakefield.

In Figure 8, we give the stopping power per pulse ion when beam density $\rho_{b0}=2.0 \times 10^{18} m^{-3}$. Results for pulses in isotropic and magnetized plasmas are shown for comparison. One can see that for high-density pulses, magnetic field reduces the stopping power for all beam velocities. This behavior occurs because the magnetic field reduces the screening of the ion-beam pulses. As the screening effect of the electrons weakens, the collective effects of the ion-beam pulses become more significant. This enhances the electron trapping processes and consequently decreases the stopping power. Moreover,  for low-velocity pulse when  $V_{b0}=0.025c$, the whistler wave excitations are inhibited, and this leads to the reduction of the stopping power, either.

\section{\label{sec:level4}CONCLUSIONS}

In this paper, we use 2D3V PIC simulations to study the wakefield and stopping power of an ion-beam pulse moving in magnetized plasmas.
Beams moving in plasmas without magnetic field are also studied for comparison.
The influence of beam velocity and density on the wake and stopping power is investigated.

For beam pulses with low beam density (the beam densities are smaller than the plasma electron density), the wakes induced by low-velocity  pulses exhibit conversed V-shape structures. Beside, the strong whistler waves are observed on both sides of the beam trajectory. The corresponding stopping power is enhanced due to the drag of these whistler waves. As beam velocities increase, the whistler waves disappear, and only are conversed V-shape wakes observed. The corresponding stopping powers decrease compared with these in isotropic plasmas.

For high-density beam pulses (the beam densities are much larger than the plasma electron densities), the collective effect dominates. The whistler wave excitations are inhibited for low-density pulses. As the velocity increases, the whistler waves disappear, and only are conversed V-shape wakes observed. The presence of the magnetic field will enhances the collective effect as well as the trapping process, and consequently reduces the stopping power for all beam velocities.

\section{Acknowledgement}

This work is supported by the National Natural Science Foundation of
P. R. China (Grant No.11505261,U1532263,11275241) and the National Magnetic Confinement Fusion
Science Program of China (Grant No. 2014GB104002).

\newpage
\begin{figure*}
\includegraphics{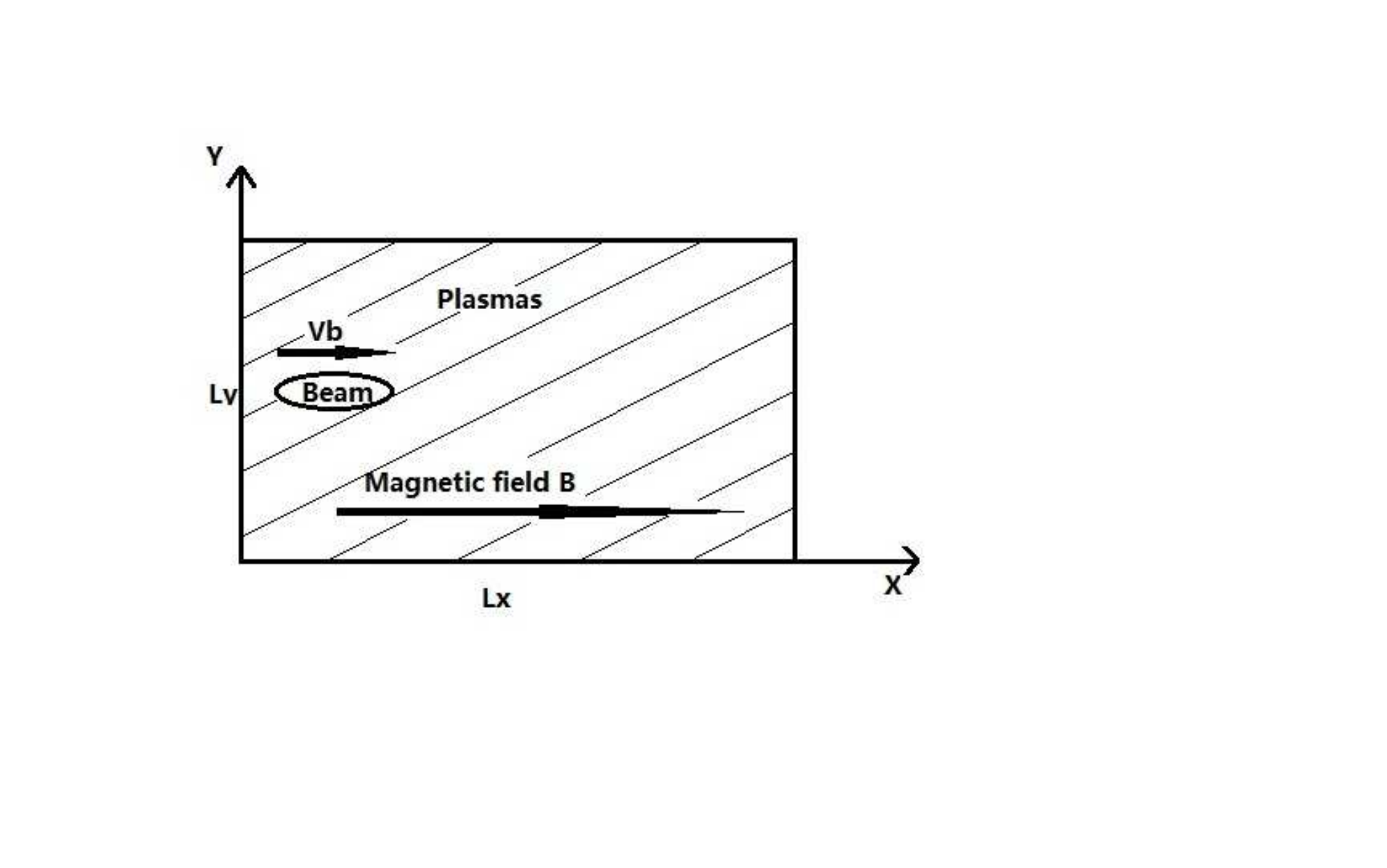}
\caption{\label{fig:model1} A two-dimensional plasma slab model.}
\end{figure*}

\newpage
\begin{figure*}
\includegraphics{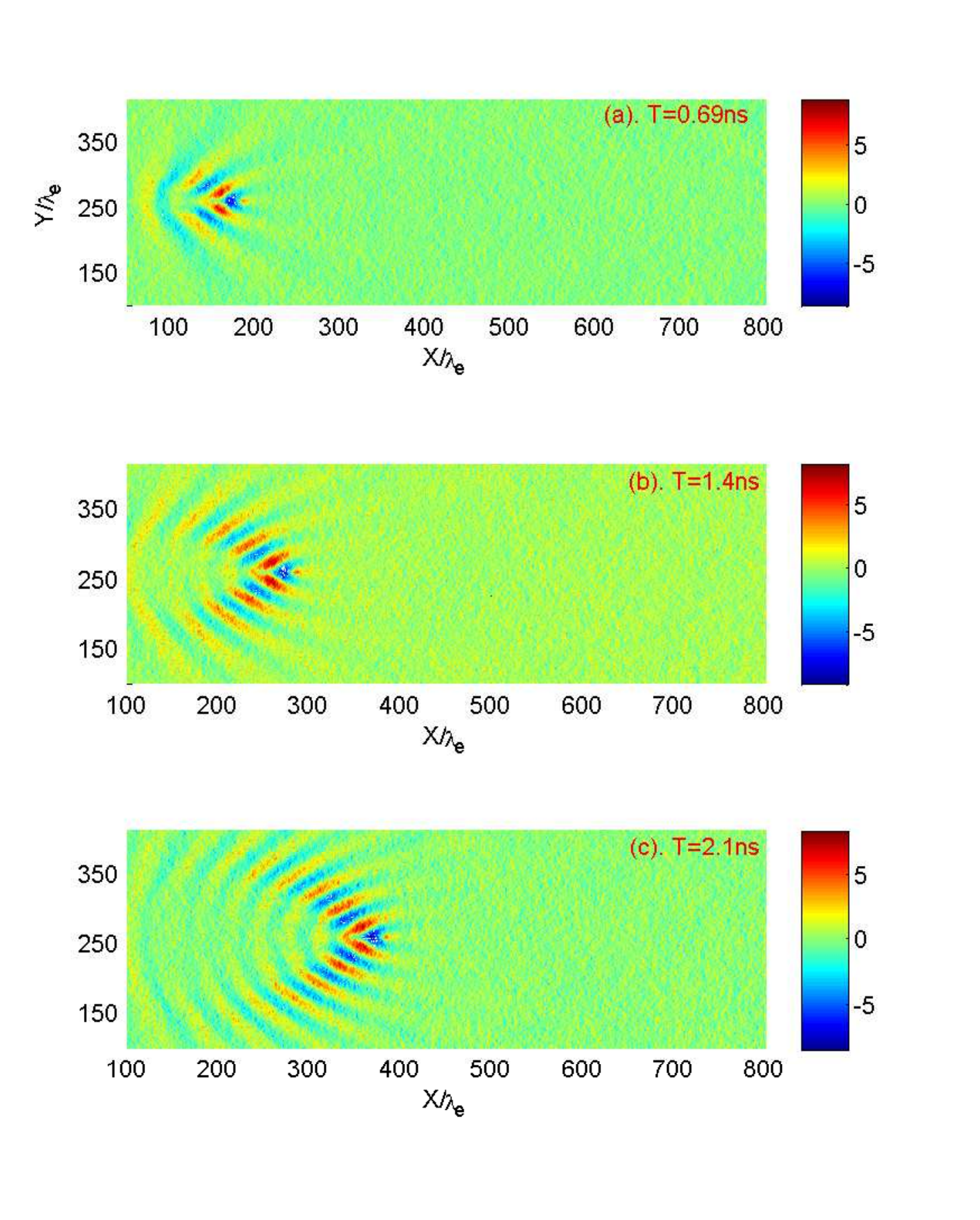}
\caption{\label{fig:model2} The time evolution of an ion-beam pulse injected
into a magnetized plasmas with magnetic field $B=1.0T$, injection
velocity $V_{b0}=0.025c$, and beam density $\rho_{b0}=1.0\times 10^{17}m^{-3}$.
The electric field $E_{ind}$ ($\times 10^{4}V/m$) induced by the pulse at three different points in time
are displayed in the figure: (a)$T=0.69ns$, (b)$T=1.4ns$, and (c)$T=2.1ns$. }
\end{figure*}

\newpage
\begin{figure*}
\includegraphics{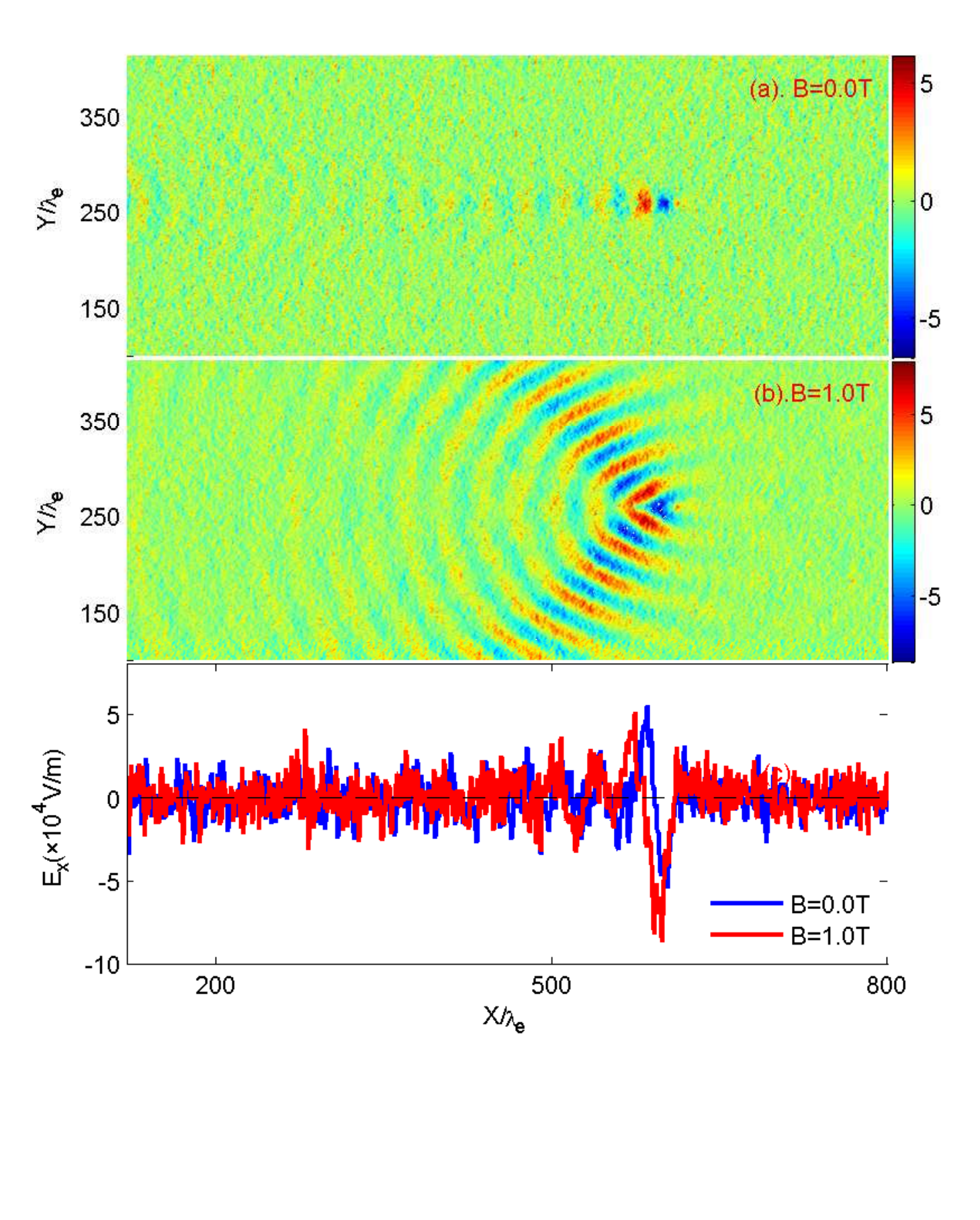}
\caption{\label{fig:model3} Contour plot of the electric field $E_{ind}$
($\times 10^{4}V/m$) in the wake field region induced by the charged particles
moving in the plasmas at the beam velocity $V_{b0}=0.025c$,
beam density $\rho_{b0}=1.0\times 10^{17}m^{-3}$,
and time $t=3.6ns$ for (a)B=0.0T and (b)B=1.0T. (c) is the corresponding longitudinal electric field along the trajectory of the pulse. The blue line is B=0.0T and the red line is B=1.0T.}
\end{figure*}

\newpage
\begin{figure*}
\includegraphics{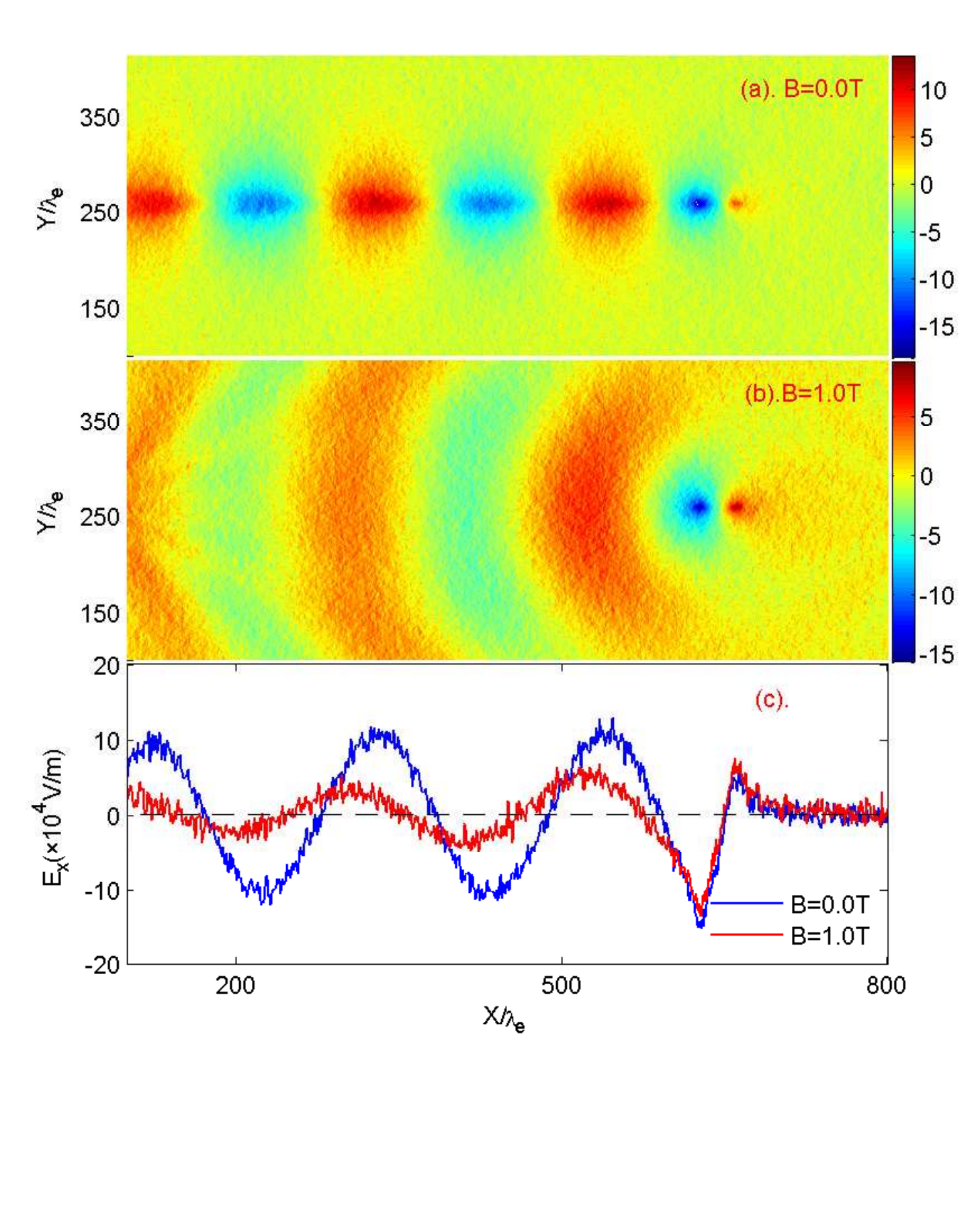}
\caption{\label{fig:model4} Contour plot of the electric field $E_{ind}$
($\times 10^{4}V/m$) in the wake field region induced by the charged particles
moving in the plasmas at the beam velocity $V_{b0}=0.15c$,
beam density $\rho_{b0}=1.0\times 10^{17}m^{-3}$,
and time $t=0.68ns$ for (a)B=0.0T and (b)B=1.0T. (c) is the corresponding longitudinal electric field  along the trajectory of the pulse. The blue line is B=0.0T and the red line is B=1.0T. }
\end{figure*}

\newpage
\begin{figure*}
\includegraphics{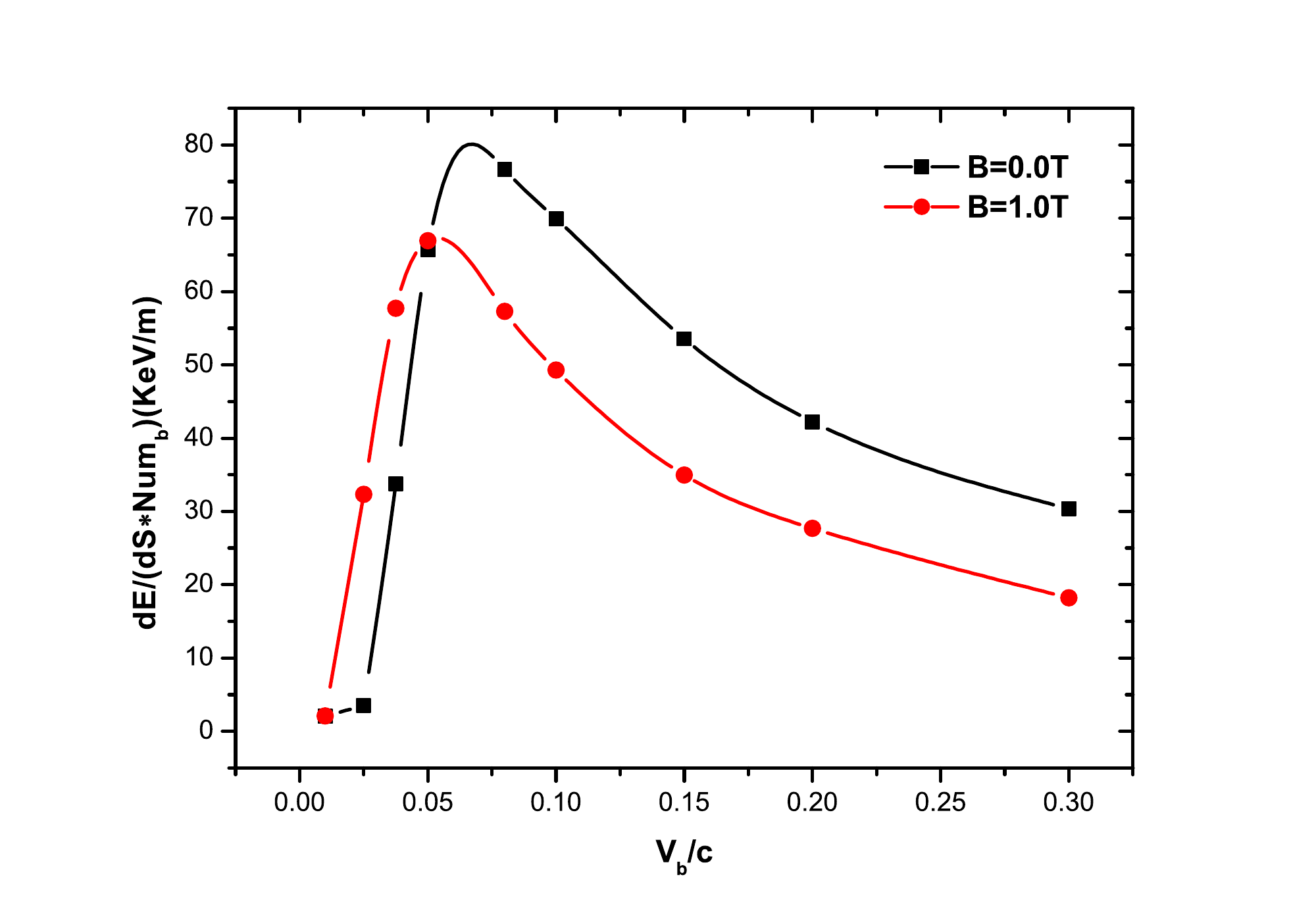}
\caption{\label{fig:model5} Stopping power per ion as a function of beam velocity
when beam density $\rho_{b0}=1.0\times 10^{17}m^{-3}$ with different magnetic fields: the black square is $B=0.0T$ and the red circle is
$B=1.0T$. The line is a linear fit of the simulation data.}
\end{figure*}

\newpage
\begin{figure*}
\includegraphics{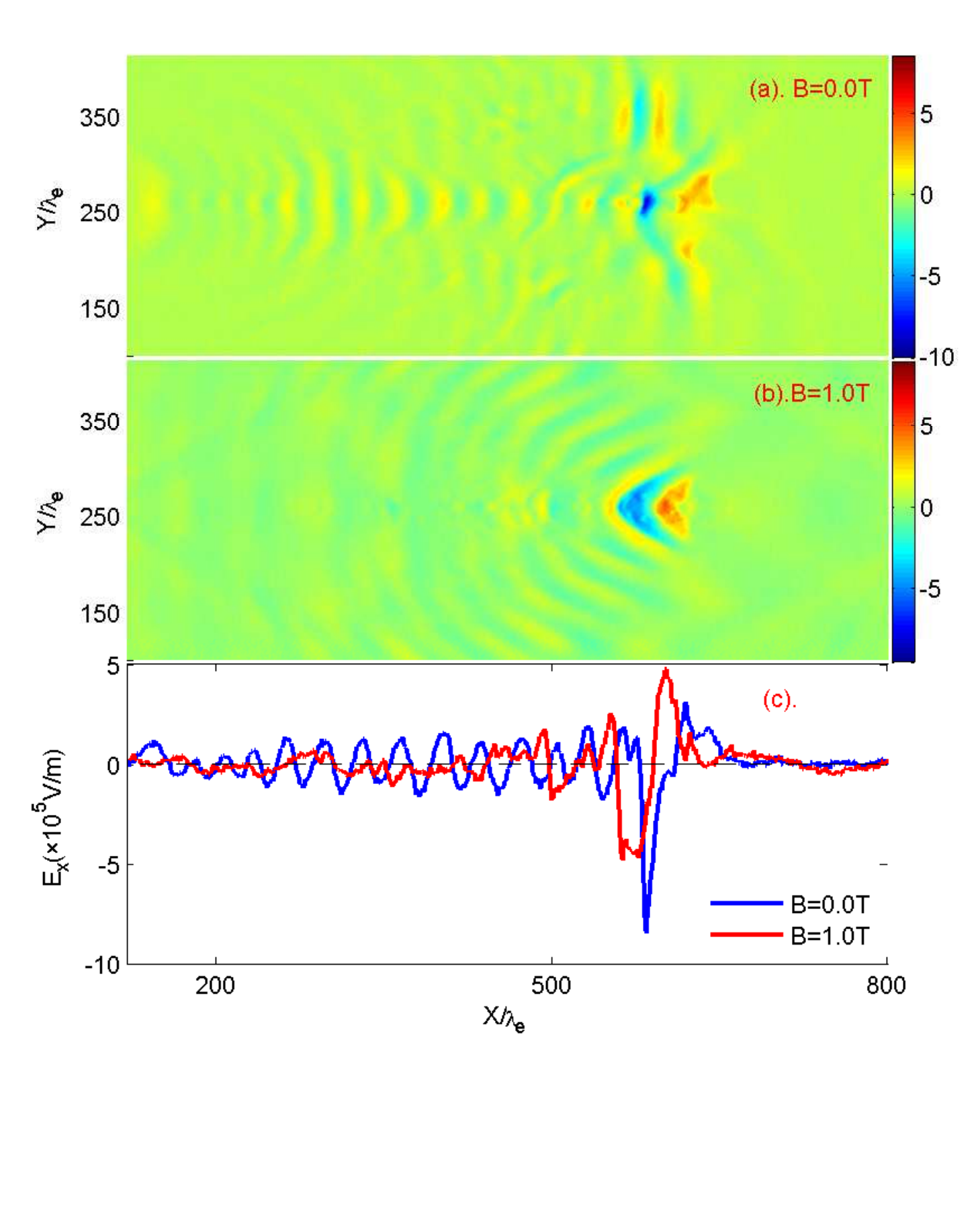}
\caption{\label{fig:model6}  Contour plot of the electric field $E_{ind}$
($\times 10^{5}V/m$) in the wake field region induced by the charged particles
moving in the plasmas at the beam velocity $V_{b0}=0.025c$,
beam density $\rho_{b0}=2.0\times 10^{18}m^{-3}$,
and time $t=3.6ns$ for (a)B=0.0T and (b)B=1.0T. (c) is the corresponding longitudinal electric field  along the trajectory of the pulse. The blue line is B=0.0T and the red line is B=1.0T.  }
\end{figure*}

\newpage
\begin{figure*}
\includegraphics{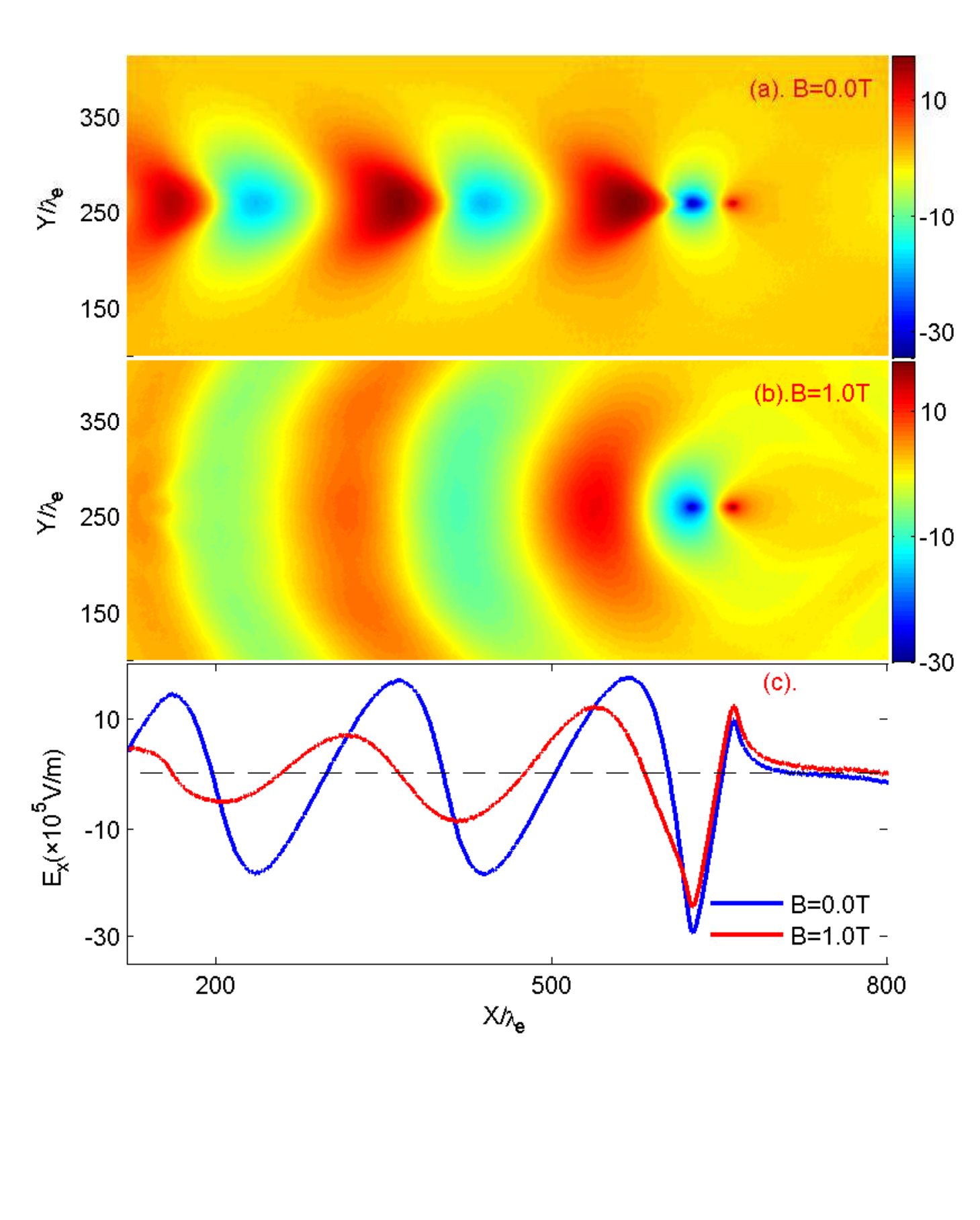}
\caption{\label{fig:model7}  Contour plot of the electric field $E_{ind}$
($\times 10^{5}V/m$) in the wake field region induced by the charged particles
moving in the plasmas at the beam velocity $V_{b0}=0.15c$,
beam density $\rho_{b0}=2.0\times 10^{18}m^{-3}$,
and time $t=0.68ns$ for (a)B=0.0T and (b)B=1.0T. (c) is the corresponding longitudinal electric field  along the trajectory of the pulse. The blue line is B=0.0T and the red line is B=1.0T. }
\end{figure*}

\newpage
\begin{figure*}
\includegraphics{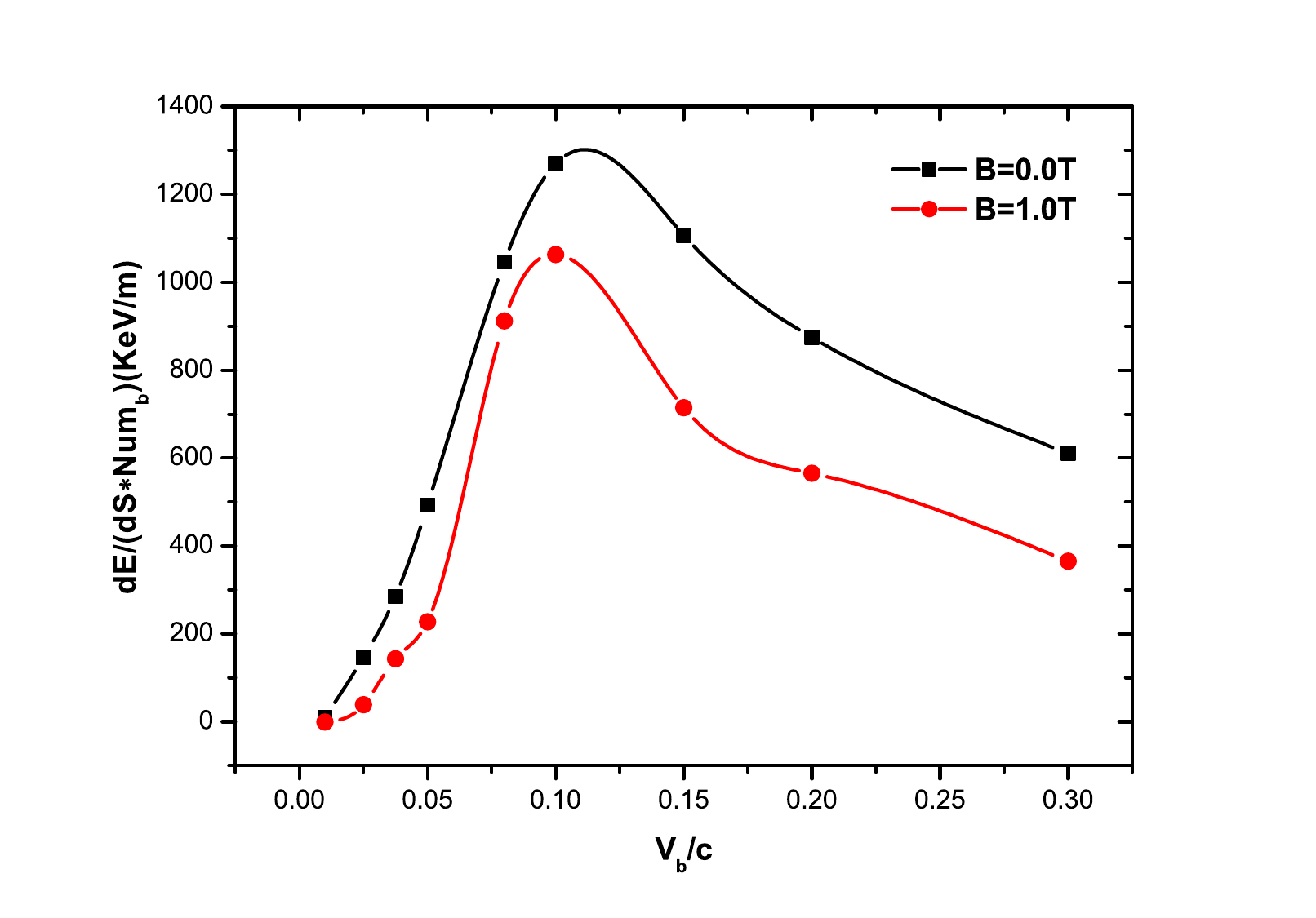}
\caption{\label{fig:model8}  Stopping power per ion as a function of beam velocity
when beam density $\rho_{b0}=2.0\times 10^{18}m^{-3}$ with different magnetic fields: the black square is $B=0.0T$ and the red circle is
$B=1.0T$. The line is a linear fit of the simulation data.}
\end{figure*}


\begin{thebibliography}{99}

\bibitem{Drake}R. P. Drake, High-Energy-Density Physics. Berlin: Springer-Verlag, (2006).
\bibitem{Sorensen1983} A. H. Sorensen and E. Bonderup, {Nucl. Instrum. Methods Phys. Res.} \textbf{215}, 27(1983).
\bibitem{Goldman1990}S. R. Goldman and I. Hofmann, {IEEE Trans. Plasma Sci.} \textbf{18}, 789(1990).
\bibitem{Renk2008}T. J. Renk, G. A. Mann, and G. A. Torres, {Laser Part. Beams} \textbf{26}, 545(2008).
\bibitem{Ter2008}S. T. Avetisyan, M. Schnuerer, R. Polster, P. V. Nickles, and W. Sandner, {Laser Part. Beams} \textbf{26}, 637(2008).
\bibitem{Thompson1993}E. Thompson, D. Stork, and H. P. L. Deesch, {Phys. Fluids B} \textbf{5}, 2468(1993).
\bibitem{Takahashi2004} T. Takahashi, T. Kato, Y. Kondoh, and N. Iwasawa, {Phys. Plasmas} \textbf{11}, 3801(2004).
\bibitem{wake1}L. -Y. Zhang, X. -Y. Zhao, X. Qi, W. -S. Duan, G. -Q. Xiao, and L. Yang, {Phys. Plasmas} \textbf{22}, 053109(2015).
\bibitem{wake2}X. -Y. Zhao, L. -Y. Zhang, Y. -L. Zhang, W. -S. Duan, X. Qi, L. Yang, and J. Shi, {Phys. Plasmas} \textbf{22}, 093114(2015).
\bibitem{wake3}O. B. Frankenheim, E. Gjonaj, F. Petrov, F. Yaman, T. Weiland, and G. Rumolo, {Phys. Rev. ST} \textbf{15}, 054402(2012).
\bibitem{wake4}I. D. Kaganovich, E. A. Startsev, and R. C. Davidson, {Phys. Plasmas} \textbf{11} 3546(2004).
\bibitem{wake5}Z. -H. Hu, Y. -H. Song, and Y. -N. Wang, {Phys. Rev. E} \textbf{82}, 026404(2010).
\bibitem{wake6}I. D. Kaganovich, G.Shvets, E. Startsev, and R. C. Davidson, {Phys. Plasmas} \textbf{8}, 4180(2001).
\bibitem{acceler}G. Sharma, G. Mishra, and Y. C. Huang, {Nucl. Instrum. Methods Phys. Res. A} \textbf{648}, 22(2011).
\bibitem{Caldwell2009}A. Caldwell, K. Lotov, A. Pukhov, and F. Simon, {Nat. Phys.} \textbf{5}, 363(2009).
\bibitem{Milhail2010} M. A. Dorf, I. D. Kaganovich, E. A. Startsev, and R. C. Davidson, Phys. Plasmas, {\bf 17}, 023013(2010)
\bibitem{Winkler1996}T. Winkler, K. Beckert, F. Bosch, H. Eickhoff, B. Franzke, O. Klepper, F. Nolden,
H. Reich, B. Schlitt, P. Sp$\ddot{a}$dtke, and M. Steck, {Hyperfine Interact.} \textbf{99}, 277(1996).
\bibitem{Boine1996} O. B. Frankenheim and J. D'Avanzo, {Phys. Plasmas} \textbf{3}, 792(1996).
\bibitem{Zwicknagel1998}G. Zwicknagel, {Nucl. Instrum. Methods Phys. Res. A} \textbf{415}, 680(1998).
\bibitem{Gericke1999}D. O. Gericke and M. Schlanges, {Phys. Rev. E} \textbf{60}, 904(1999).
\bibitem{Hu2009}Z. -H. Hu, Y. -H. Song, G. -Q. Wang, and Y. -N. Wang, {Phys. Plasmas} \textbf{16}, 112304(2009).
\bibitem{Zwicknagel1996}G. Zwicknagel, P.-G. Reinhard, C. Seele, and C. Toepffer, {Fusion Engineering and Design} \textbf{32-33}, 523(1996).
\bibitem{Bringa1995}E. M. Bringa and N. R. Arista, {Phys. Rev. E} \textbf{52}, 3010(1995).
\bibitem{Walter2000}M. Walter, C. Toepffer, and G. Zwicknagel, {Nucl. Instrum. Methods Phys. Res. B} \textbf{168},347(2000).
\bibitem{Mollers2003}B. M\"{o}llers, M. Walter, G. Zwicknagel, C. Carli, C. Toepffer, {Nucl. Instrum. Methods Phys. Res. B} \textbf{207},462(2003).
\bibitem{collective1}J. D'Avanzo, M. Lontano, and P. F. Bortiggnon, Phys. Rev. A, \textbf{45}, 6126(1992).
\bibitem{collective2}J. D'Avanzo, M. Lontano, and P. F. Bortiggnon, Phys. Rev. E, \textbf{47}, 3574(1993).

\bibitem{vorpal}C. Nieter and J. R. Cary, {J. Comput. Phys.} \textbf{196}, 448(2004).

\end{thebibliography}
\end{document}